\documentclass{aa}  
\usepackage[dvipsnames]{xcolor}
\usepackage{graphicx}
\usepackage{txfonts}

\usepackage[colorlinks=true, linkcolor=blue, urlcolor=blue, citecolor=blue]{hyperref}
\makeatletter
\renewcommand*\aa@pageof{, page \thepage{} of \pageref*{LastPage}}
\makeatother
\usepackage{xspace}

\def\geff{\ensuremath{\gamma_{\textrm{eff}}}\xspace}

\begin{document}

   \title{Dust back-reaction on gas around planets modifies the cold thermal torque}

  \author{Ra\'ul~O.~Chametla\inst{\ref{UKarlova}}
          \and
          Ond\v{r}ej~Chrenko\inst{\ref{UKarlova}}
          \and
          F.~J.~S\'anchez-Salcedo\inst{\ref{UNAM}}
          \and
          Mauricio~Reyes-Ruiz\inst{\ref{UNAM-Ensenada}}
          \and
          Cl\'ement~Baruteau\inst{\ref{IRAP}}
          \and 
          Alicia~Moranchel-Basurto\inst{\ref{UKarlova}}
          \and
          Joanna~Dr{\k{a}}{\.z}kowska\inst{\ref{MPS}}
          \and
          Gabriel-Dominique~Marleau\inst{\ref{Bern},\ref{MPIA},\ref{UDE}}
          \and
          Yasuhiro~Hasegawa\inst{\ref{JPL}}
          \and
          Sonia~Cornejo\inst{\ref{MPS}}
          }
    \authorrunning{R.~O.~Chametla et al.}
   \institute{%
            Charles University, Faculty of Mathematics and Physics, Astronomical Institute,
            V Hole\v{s}ovi\v{c}k\'ach 747/2, 180 00,
            Prague 8, Czech Republic\\
            \email{raul@sirrah.troja.mff.cuni.cz}
             \label{UKarlova}
         \and
            Instituto de Astronomía, Universidad Nacional Autónoma de México, %Ciudad Universitaria, 
            Apt.~Postal 70-264, C.P.~04510, Mexico City, Mexico
             \label{UNAM}
        \and 
            Instituto de Astronomía, Universidad Nacional Autónoma de México, Ensenada, 22800 B.C., Mexico
             \label{UNAM-Ensenada}
        \and
            IRAP, Universit\'e de Toulouse, CNRS, UPS, F-31400 Toulouse, France
            \label{IRAP}
        \and
            Max-Planck-Institut f\"ur Sonnensystemforschung, Justus-von-Liebig-Weg 3, G\"ottingen 37077, Germany
                \label{MPS}
        \and
             Division of Space Research \&\ Planetary Sciences, Physics Institute, University of Bern, Sidlerstr.~5, 3012 Bern, Switzerland%
            \label{Bern}
        \and
            Max-Planck-Institut f\"ur Astronomie,
            K\"onigstuhl 17,
            69117 Heidelberg, Germany
            \label{MPIA}
        \and
            Fakult\"at f\"ur Physik,
            Universit\"at Duisburg-Essen,
            Lotharstra\ss{}e 1,
            47057 Duisburg, Germany
            \label{UDE}
        \and
            Jet Propulsion Laboratory, California Institute of Technology, Pasadena, CA 91109, USA
            \label{JPL}
        }

   \date{Received XXX; accepted YYY}

% \abstract{}{}{}{}{} 
% 5 {} token are mandatory
 
  \abstract
   {A nascent planet in a gas disk experiences radial migration due to the different torques which act on it (e.g., Lindblad and corotation torques). It has recently been shown that the torques produced by the gas and dust density variations around a non-accreting low-mass planet, the so-called cold thermal and dust streaming torques, can surpass each of the other torque components.}
   {We investigate how the total torque acting on the planet is affected by the presence of dust grains and their aerodynamic back-reaction on gas, while taking into account the cold thermal torque produced by thermal diffusion in the gas component.}
   {We perform high-resolution local and global three-dimensional two-fluid simulations within the pressureless-fluid dust approximation using the \textsc{Fargo3D} code. We explore the influence of different dust species parameterized by the Stokes number, focusing on non-accreting protoplanets with masses from one-third the mass of Mars to one Earth mass.}
   {The dust feedback has substantial impact on the asymmetry of the cold thermal lobes (which produce the cold thermal torque). However, the total torque is dominated by the dust torque when $\mathrm{St}>10^{-2}$. The dust torque becomes more negative over time due to the formation of dust lobes that resemble the cold thermal lobes that form in the gas component. Therefore, the dust streaming torque prevails over the cold thermal torque. On the other hand, when $\mathrm{St}\leq10^{-2}$, the dust streaming torque is negligible and thus, the total torque on the planet comes from the gaseous component of the disk.
   }
   {Our results suggest that a planet embedded in a gas-dust disk may experience stagnant migration or inward runaway migration in regions of the protoplanetary disk where the dust is not fully coupled to the gas. However, this behaviour could change in regions with strong dust--gas coupling or in the inner transition region of the disk, where the cold thermal torque may become relevant.
   }


   \maketitle
%
%-------------------------------------------------------------------

\section{Introduction}

Planets are formed in disks of gas and dust and, as their mass grows, they interact gravitationally with the disk. The exchange of angular momentum between the planet and the disk due to the gravitational interaction substantially changes the planet's semi-major, which is known as the process of planetary migration (see \citealt{Paar_etal2023} for a review). For isothermal disks with simple power-law profiles, the disk's torque leads to a decrease of the planet's orbital radius. A long outstanding problem is that
low-mass planets would fall toward the central star 
in a timescale much shorter than the lifetime of the disk
\citep[][]{Ward1986,TTW2002}. In the last two decades, different mechanisms have been proposed to solve this problem. They include non-isothermal disks, disks with a magnetic field, vortex--planet interactions, density traps, energy transfer within the disk that gives rise to ``thermal torques'' (defined hereafter), the dust torques (which arise from the asymmetric distribution of dust around the planet), among others \citep[see][for a review]{Lesur_etal2022,Paar_etal2023}. 

%#########################
\begin{table}
\caption{\label{tab:t1}Summary of our numerical models. \textit{First column.} Model name, here the capital letters in model names have the following meaning. LR: linear regime, AD: adiabatic disk, TD: thermal diffusivity. \textit{Second column.} The mass of the planet (in units of Earth masses). \textit{Third column}. Dimensionless Stokes number (see Eq. \ref{eq:St}). \textit{Fourth column}. Flag of the inclusion of the back-reaction force from the dust on gas. Note that models with feedback and the same value of the planet's mass represent a two-fluid simulation.}
\centering
\resizebox{9cm}{!}{
\begin{tabular}{lccc}
\hline\hline
Name&Planet Mass&Stokes Number&Feedback\\
&$[M_\oplus]$&(St)&\\
\hline
Mp0.03LR\tablefootmark{a}&0.03&0.0 &No\\
&&$2.8\times10^{-5}$ &No\\
&&$2.8\times10^{-5}$ &Yes\\
&&$0.0028$ &Yes\\
&&0.028 &Yes\\
&&0.28 &Yes\\

\hline
%Mp0.3AD\tablefootmark{a}&0.3&0.28 &Yes\\
Mp0.3AD&0.3&0.0 &No\\
&&0.001 &Yes\\
&&0.1 &Yes\\
Mp0.3TD&0.3&0.001 &Yes\\
&&0.1 &Yes\\
\hline
Mp1.0AD&1.0&0.001 &Yes\\
&&0.01 &Yes\\
&&0.1 &Yes\\
Mp1.0TD&1.0&0.001 &Yes\\
&&0.01 &Yes\\
&&0.1 &Yes\\
\hline

\end{tabular}}
\tablefoot{
\tablefoottext{a}{Model in the local mesh approach.}
}
\end{table}

%#########################

Recent work has shown that thermal diffusion in gaseous disks can play a decisive role in determining the direction of migration \citep[][]{Lega_etal2014,BLl_etal2015,Masset2017} and that the total torque can differ greatly from that in isothermal or adiabatic disks \citep[][]{JM2017,ChM2021}. In the case of a non-luminous planet embedded in a gaseous disk with thermal diffusion, the torque difference with respect to the adiabatic case is called the cold thermal torque. This torque arises from the gas in the vicinity of the planet which becomes colder and denser as the energy generated by compressional heating diffuses away from the planet. The distribution of cold gas around the planet develops a shape of two symmetric lobes if the disk is strictly Keplerian. When the planet is offset from the planet--gas corotation, as occurs when there is a radial pressure gradient, the lobes are not symmetric with respect to the planet. The lobe located on the same side of corotation as the planet is more pronounced, resulting in a net torque on the planet with the same sign as the torque from this lobe \citep[][]{Masset2017}. \citet{Lega_etal2014} showed that in the usual case in which the disk is slightly sub-Keplerian and the planet is outside of its corotation, the cold thermal torque is negative, that is, the migration is inward. The effect of thermal diffusion on the gas density distribution around a low-mass planet was first reported in \citet[][]{Lega_etal2014}. Using 3D radiative hydrodynamical simulations, they found two cold regions of high density in the co-orbital region near the planet. These regions, which were named "cold fingers" by the authors, are the result of a larger entropy loss in gas parcels that experience more compressional heating.

Later on, \citet{Masset2017} showed that these structures in density are mainly due to the fact that thermal diffusion flattens the temperature peak that is located at the planet position.  Additionally, he suggested that the denomination "cold lobes" is more adequate in the presence of thermal diffusion and showed that 3D calculations are required to correctly capture the respective cold thermal torque.

On the other hand, given the low mass of dust in protoplanetary disks (the dust-to-gas ratio is considered to be around 0.01), the gravitational torque exerted by the dust on the planet is typically neglected. However, \citet{BLlP2018}, using a 2D isothermal gas-dust model without feedback of the dust on the gas, found that the dust in the vicinity of the planet can generate a positive total torque depending on the dust drift speed, the dust-to-gas mass ratio/distribution, and the embryo mass \citep[see also][]{Regaly2020,Chrenko2024}. 

Previous studies that included thermal diffusion did not consider the contribution of dust to the total torque. In this paper, we incorporate both thermal diffusion and the dust back-reaction on the gas (hereafter named ``feedback'') to investigate whether the cold, dense lobes are redistributed differently depending on the degree of dust-gas coupling, and how this affects the total torque acting on the planet.

The paper is organized as follows. In Section~\ref{sec:init} we present the physical model. In Section~\ref{sec:setup}, the numerical setup used in our high-resolution 3D simulations is described. We show the main effects of the dust feedback on the cold thermal torque and on the resulting total torque in Section~\ref{sec:results}. In Section~\ref{sec:discussion} we discuss the main implications of the behavior of the total torque on the migration of a low-mass planet. Concluding remarks can be found in Section~\ref{sec:conclusions}.
%--------------------------------------------------------------------

\section{Physical model}
\label{sec:init}

In this section, we describe the different components of our physical model and provide a brief description of 3D gaseous-dusty disks employed in this study.

\subsection{Gas disk governing equations}
\label{subsec:gas}
We consider a 3D, non-self-gravitating gas disk component whose evolution is governed by the following equations:
\begin{align}
\partial_t\rho_\mathrm{g}+\nabla\cdot(\rho_\mathrm{g} \mathbf{v}) &=0,  \label{eq:gas_cont}\\
\partial_t(\rho_\mathrm{g}\mathbf{v})+\nabla\cdot(\rho_\mathrm{g}\mathbf{v}\otimes\mathbf{v}+p\mathsf{I}) &= -\rho_\mathrm{g}\nabla\Phi+\nabla\cdot\mathrm{\tau}-\mathbf{f}_\mathrm{d},  \label{eq:gas_mom}\\
\partial_te_\mathrm{g}+\nabla\cdot(e_\mathrm{g}\mathbf{v}) &=-p\nabla\cdot\mathrm{v}-\nabla\cdot\mathbf{F}_\mathrm{H},  \label{eq:gas_energy}
\end{align}
where $\rho_\mathrm{g}$ and $\mathbf{v}$ denote the gas density and the gas velocity, respectively. In Eq.~(\ref{eq:gas_energy}), $\mathbf{F}_\mathrm{H}$ is the heat flux, given as
\begin{equation}
\mathbf{F}_\mathrm{H}=-\chi\rho_\mathrm{g}\nabla\left(\dfrac{e_\mathrm{g}}{\rho_\mathrm{g}}\right),
    \label{eq:heatflux}
\end{equation}
where $e_\mathrm{g}$ is the gas thermal energy density and $\chi$ the thermal diffusivity. Furthermore, $\Phi$ denotes the gravitational
potential, $\mathsf{I}$ is the unit tensor, $\mathbf{f}_\mathrm{d}$ is the drag force between the gas and dust (Eq.~\ref{eq:dragforce}), and $p$ is the gas pressure. For the latter, we write the caloric equation of state as
\begin{equation}
p=\left(\geff-1\right)e_\mathrm{g},
 \label{eq:pressure}
\end{equation}
and use $\geff=7/5$, the value for a perfect diatomic gas. This applies approximately to a solar-abundance H$_2$+He gas above $T\approx300$~K \citep{BB2013} and below $T\approx1000$--2000~K, as one can verify easily for instance using the \citet{Vetal2015} implementation of the classical equation of state formul\ae{} of \citet{DAB13}. (The latter use $\Upsilon$, an uppercase Upsilon, instead of \geff.)
In this low-temperature regime, the first adiabatic exponent $\Gamma_1$ (not to be confused with a torque) is also approximately equal to 7/5. It will be used in Equation~(\ref{eq:St}) below. Afterwards, we will denote both indistinctly by $\gamma$ but for a more general (even ideal but not constant) equation of state, they would need to be distinguished.

In our global 3D disk models, the gas turbulence is included in Eq. (\ref{eq:gas_mom}) via the viscous stress tensor:
\begin{equation}
\mathbf{\tau}=\rho_\mathrm{g}\nu\left[\nabla\mathbf{v}+(\nabla\mathbf{v})^{\mathrm{T}}-\frac{2}{3}(\nabla\cdot\mathbf{v})\mathsf{I}\right],
    \label{eq:visc}
\end{equation}
where $\nu$ is the gas kinematic viscosity parametrised as in $\alpha$ description \citep[see][]{SS1973}.
 
We stress that gas viscosity has an important implicit effect on dust dynamics, since through it, dust itself feels turbulence.

Lastly, we mention that our local models do not include the gas turbulence, i.e. we use an inviscid disk model.

\subsection{The governing equations for the dust}
\label{subsec:dust}
The equations of dust dynamics in the pressureless fluid approximation \citep{Webber_etal2018} are given as
\begin{align}
\partial_t\rho_\mathrm{d}+\nabla\cdot(\rho_\mathrm{d} \mathbf{u}+\mathbf{j}_\mathrm{d}
) &=0,  \label{eq:dust_cont}\\
\partial_t(\rho_\mathrm{d}\mathbf{u})+\nabla\cdot(\rho_\mathrm{d}\mathbf{u}\otimes\mathbf{u}) &=-\rho_\mathrm{d}\nabla\Phi+\mathbf{f}_\mathrm{d}.
 \label{eq:dust_mom}
\end{align}
Here $\rho_\mathrm{d}$ and $\mathbf{u}$ are the dust density and dust velocity, respectively. In Equations~(\ref{eq:gas_mom}) and (\ref{eq:dust_mom}), the last term on the right-hand side is the drag force term given as
\begin{equation}
    \mathbf{f}_\mathrm{d}=\frac{\Omega }{\mathrm{St}}\rho_\mathrm{d}(\mathbf{v}-\mathbf{u}),
	\label{eq:dragforce}
\end{equation}
where $\mathrm{St}$ is the Stokes number, which measures the dust–gas coupling and relates to the dust grain properties via
\begin{equation}
    \mathrm{St}=\sqrt{\frac{\pi\Gamma_1}{8}}\frac{a\rho_\bullet\Omega}{\rho_\mathrm{g}c_\mathrm{s}}=\sqrt{\frac{\pi}{8}}\frac{a\rho_\bullet\Omega}{\sqrt{p\rho_\mathrm{g}}},
    \label{eq:St}
\end{equation}
where $a$ is the grain size, $c_\mathrm{s}^2 = \Gamma_1 p/\rho_\mathrm{g}$ is the sound speed (squared), with $\Gamma_1$ the adiabatic index, and $\rho_\bullet$ the material density of the dust.

\subsection{Turbulent dust diffusion model}

Since gas turbulence has been included in global simulations, its effect on the dust component must also be
considered. Therefore, we incorporate this effect into Eq. (\ref{eq:dust_cont}) through the vector $\mathbf{j}_\mathrm{d}$ \citep[][]{MorfilV1984}:
\begin{equation}
\mathbf{j}_\mathrm{d}=-D_\mathrm{d}(\rho_\mathrm{g}+\rho_\mathrm{d})\nabla\left(\frac{\rho_\mathrm{d}}{\rho_\mathrm{g}+\rho_\mathrm{d}}\right),
    \label{eq:dd}
\end{equation}
where $D_\mathrm{d}$ is the dust diffusion coefficient, which is linked to gas turbulence through the Schmidt number
\begin{equation}
\mathrm{Sc}=\frac{\nu}{D_\mathrm{d}},
    \label{eq:Sch}
\end{equation}
which quantifies the relative effectiveness of the gas angular momentum transport and the dust-mixing processes. It is common to assume that the dust diffusion turbulent coefficient $\delta\equiv D_\mathrm{d}/(c_sH_\mathrm{g})$, with $H_\mathrm{g}$ is the pressure scale height of the gas in the disk, defined as $H_\mathrm{g}=c_s/\Omega_\mathrm{Kep}$ (here $\Omega_\mathrm{Kep}$ the Keplerian angular frequency), is of the order of the $\alpha$ viscosity parameter. Nevertheless, we are aware that $\alpha$ does not need to be the same as $\delta$ \citep{Krapp2024}. Since the difference arises from that while $\alpha$ is set by ionized gas dynamics in hot gas regions or by hydrodynamical instabilities such as vertical shear instability \citep[VSI;][]{Lesur2025,Ogil2025} or spiral wave perturbations \citep[][]{CH2024} in cold gas regions (which applies here), $\delta$ is controlled by gas-dust interactions and the size of turbulent eddy \citep[see][]{Zhu2015}. Furthermore, the turbulent alpha coefficient in the gas is always larger than the turbulent diffusion coefficient of the dust \citep[see for instance][and references therein]{Hasegawa2017}.

Note that here, we have followed a similar approach that of \citet{CH2025} where $\nu= D_\mathrm{d}$ (here the diffusion coefficients are constant). Our motivation for this choice is based on the fact that in a turbulent disk ($\alpha\simeq10^{-3}$) dust asymmetries around of Earth-like planet are not formed. This allows us to isolate the effect of dust feedback on the cold lobes and vice versa.

In the low-turbulence approximation ($\alpha=10^{-5}$) the dust hole can form in the vicinity of the planet (see Appendix \ref{ap:appendixA}) which is in agreement with what was reported in \citet{CH2025}.

\subsection{Gravitational potential}

The gravitational potential $\Phi$ is given by
\begin{equation}
\Phi=\Phi_S+\Phi_p,
 \label{eq:potential}
\end{equation}
where the stellar and planetary potentials are respectively
\begin{align}
\Phi_S &=-\frac{GM_\star}{r},  \label{eq:Star_potential}\\
\Phi_p &=-\frac{GM_p}{\sqrt{r'^2+r_\mathrm{sm}^2}}+\frac{GM_pr\cos\phi\sin\theta}{r_p^2}. \label{eq:Planet_potential}
\end{align}
In Eq.~(\ref{eq:Planet_potential}), $r'=\left|\mathbf{r}-\mathbf{r}_p\right|$ is
the cell-planet distance,
$\phi$ is the azimuth with respect to the planet, $\theta$ is the colatitude, 
and $r_\mathrm{sm}$ is a softening length used to avoid computational divergence of the potential in the vicinity of the planet. The second
term on the right-hand side of Eq.~(\ref{eq:Planet_potential}) is the so-called
``indirect term'' arising from the reflex motion of the star \citep[][]{MD1999,Muller2012,Cri2022}. 

For the smoothing length, we explored two cases: $r_\mathrm{sm} = 0.005H_\mathrm{g}$ and $r_\mathrm{sm} = 0.01H_\mathrm{g}$. Note that the smoothing length is a numerical parameter and that its value has been chosen to be sufficiently small,
on the order of a grid cell, to minimize its impact on the global solution (see Appendix \ref{ap:appendixB}).

\section{Set-up}
\label{sec:setup}
We consider a 3D gaseous-dusty disk co-rotating with the planet and use a polar spherical grid $(r,\theta,\phi)$ centered on the star. Here, $r$ is the radial distance from the star, $\theta$ is the polar angle and $\phi$ is the azimuthal angle. The volumetric gas disk density is given by \citep{MBLl2016}:
\begin{equation}
\rho_\mathrm{g}(r,\theta)=\rho^\mathrm{eq}_\mathrm{g}\,\left(\sin\theta\right)^{-\beta-\xi+h^{-2}_\mathrm{g}}
    \label{eq:rhog}
\end{equation}
with 
\begin{equation}
\rho^\mathrm{eq}_\mathrm{g}=\frac{\Sigma_0}{\sqrt{2\pi}h_\mathrm{g} r_p}\left(\frac{r}{r_p}\right)^{-\xi},
    \label{eq:rhoeq}
\end{equation}
where $\Sigma_0$ is the surface density at $r=r_p$ and $h_\mathrm{g}$ the gas disk
aspect ratio. 

For the dust disk component, we consider an initial dust density profile that depends on the dust pressure scale $h_\mathrm{d}$, such that
\begin{equation}
\rho_\mathrm{d}=\rho^\mathrm{eq}_\mathrm{d}(\sin\theta)^{-\beta-\xi+h_\mathrm{d}^{-2}}
    \label{eq:rhod}
\end{equation}
and 
\begin{equation}
\rho^\mathrm{eq}_\mathrm{d}=\frac{\epsilon\Sigma_0}{\sqrt{2\pi}h_\mathrm{d}r_p}\left(\frac{r}{r_p}\right)^{-\xi}.
    \label{eq:rhodeq}
\end{equation}
In Eq. (\ref{eq:rhodeq}), the initial dust-to-gas mass ratio $\epsilon$ is such that $\Sigma_\mathrm{d}/\Sigma_\mathrm{g}=0.01$.

We initialize the gas velocity components as follows: $v_r=v_\theta=0$ and\footnote{Note that both the density and the azimuthal velocity (Eqs. \ref{eq:rhog} and \ref{eq:vphi}) are steady-state disk solutions when $c_s\equiv constant$ along curves of constant spherical radius.} 
\begin{equation}
v_\phi=\sqrt{\frac{GM_\star}{r\sin{\theta}}-\xi c_\mathrm{s}^2.}
 \label{eq:vphi}
\end{equation} 
The radial and vertical components of the dust velocity are both zero initially, 
whereas the azimuthal component starts with the gas azimuthal rotation profile. We performed our simulations in pairs, that is, including only the gas, or the gas and a single dust species. 

%%%%%%%%%%%%%%---------
\begin{figure}
\includegraphics[width=0.4853\textwidth]{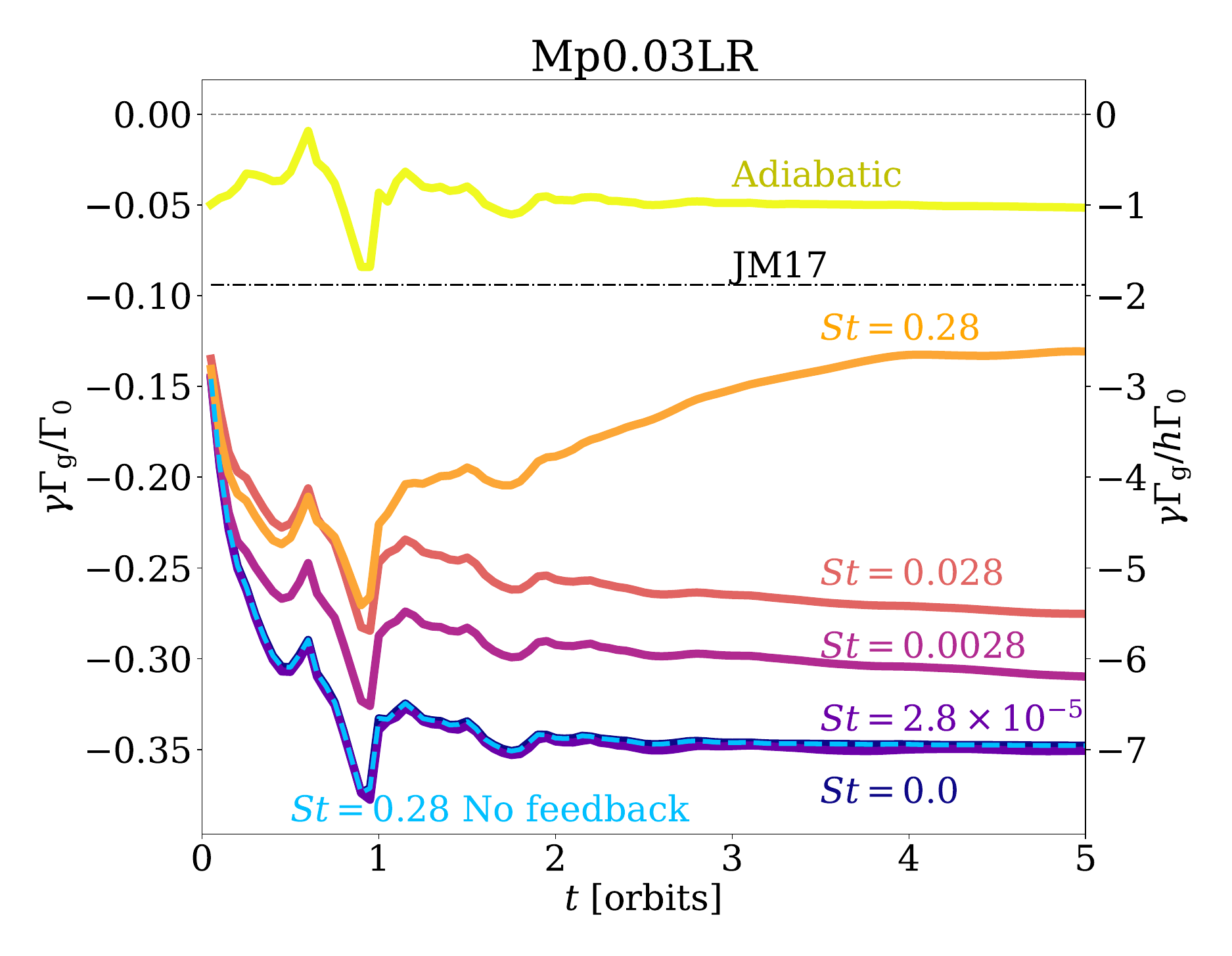}
 \caption{Normalized gas torque as a function of time in the linear case ($x_p/\lambda_c\approx0.57$) for different values of $\mathrm{St}$. The horizontal dotted line shows the torque value expected in the adiabatic disk according to \citet[][]{JM2017} which we have included as a reference value. The right vertical axis corresponds to the usual normalization of the torque.
}
\label{fig:torques}
\end{figure}
%%%%%%%%%%%%%%----------

\subsection{Relevant scales in cold thermal torque: size of thermal lobes}

\citet{Masset2017} found that thermal disturbances in the vicinity of a low-mass planet can be described by an advection-diffusion equation. Even more so, the size of the thermal disturbance can be given as 
\begin{equation}
    \label{eq:Lc}
\lambda_c=\sqrt{\frac{\chi}{(3/2)\Omega_p\gamma}},
\end{equation}
where $\Omega_p$ the planet's orbital angular frequency.
The degree of symmetry of the disturbance depends on the offset $x_p$ of the planet with respect to the corotation radius, defined as
\begin{equation}
    \label{eq:xp}
x_p=\eta h_{\rm g}^2r_p.
\end{equation}
Here, $\eta=\sigma/3+(\beta+3)/6$, with $\sigma\equiv -d\ln\Sigma/d\ln r$ and
$\beta\equiv -d\ln T/d\ln r$.

\subsection{Local inviscid disk models}
\label{subsec:local}

We will analyze the effect of dust feedback on the gas and the resulting modification of the cold thermal torque, applying it to the same model previously studied by \citet[][]{ChM2021}, who did not account for dust in their analysis. The model consists of a low-mass planet with $M_p=0.03M_\oplus$, embedded in an inviscid disk ($\nu=0$) with parameters $(\sigma,\beta)=(0,0)$, and $\chi=10^{-5}r_p^2\Omega_p$. In this model $x_p/\lambda_c\approx 0.57$, representing a case where
there is good agreement between linear theory and cold thermal torque simulations.

We include the dust component and consider four values of the Stokes number in
the range $[2.8\times 10^{-5},0.28]$. The dust density was initialized as $\rho_\mathrm{d}=\epsilon\rho_\mathrm{g}$, implying that $h_{\rm d}=h_{\rm g}$ (see Eqs.~\ref{eq:rhoeq}--\ref{eq:rhodeq}). At $t=0$, we set $\epsilon=0.01$.%} 

To enable a faithful comparison within the linear regime, we perform local simulations 
specifically designed to resolve the thermal disturbance $\lambda_c$ with sufficient accuracy.
In particular, resolving $\lambda_c$ with approximately ten grid cells is necessary to
properly capture the strength of the cold thermal torque. In the local simulations, the computational mesh extends
from $0.9r_{p}$ to $1.1r_{p}$ in the radial direction, from $-\pi/18$ to $\pi/18$ in
the azimuthal direction, and between $\pi/2-h_{p}$ and $\pi/2$ in colatitude. Here
$h_{p}$ is the gas aspect ratio at the position of the planet. 

\subsection{Global viscous disk models}
\label{subsec:global}
In the global models, we adopt disk parameters similar to those used in \citet{EM2017} and \citet{VRetal2022}. Specifically, we use
$(\beta,\xi)=(1,1.5)$, yielding a power-law index for
the surface density of $\sigma=0.5$, a surface gas density $\Sigma_0=6.05\times 10^{-4}$ at $r_p$ (about $200\,\mathrm{g \, cm^{-2}}$ for a solar-like central star), a constant aspect ratio $h_\mathrm{g}=0.05$, and a thermal diffusivity of $\chi=4.5\times 10^{-5}r_p^2\Omega_p$\footnote{We stress that while thermal diffusivity could drive a change in the Lindblad torque \citep[][]{Paar2011}, the values considered here in the local and global models fall within the thermal diffusivity values typical of protoplanetary disks $\sim10^{15}\mathrm{cm}^2\mathrm{s}^{-1}$ \citep[][]{Bitsch2014,Lega_etal2014,BLl_etal2015} and its effect is mostly reflected in the cold thermal torque \citep[see][]{Masset2017}.}. We also include a constant kinematic viscosity $\nu=10^{-5} r_p^{2}\Omega_p$, corresponding to the value of $\alpha= 2.5 \times 10^{-3}$. The mass of the planet takes the values $M_p=0.3M_\oplus$ and $1.0M_\oplus$.

%%%%%%%%%%%%%%---------
\begin{figure}
\includegraphics[width=0.4853\textwidth]{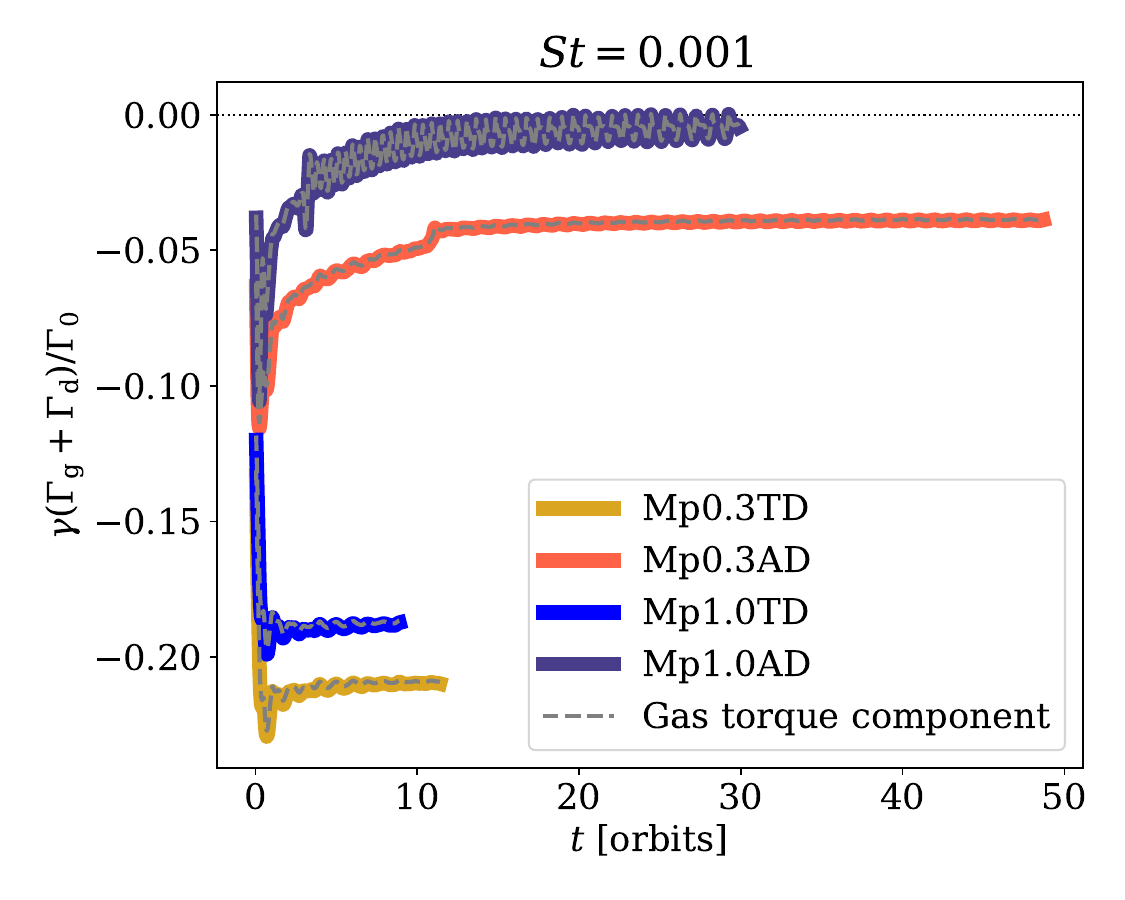}
 \caption{Normalized total torque (gas plus dust) as a function of time in the settling regime for global models (see Table \ref{tab:t1}) when $\mathrm{St}=10^{-3}$. Although $x_p/\lambda_c\approx0.45$ for all simulations, the smallest mass of the planets considered here is equal to the critical mass $M_c$ (see Eq.~\ref{eq:potential}), so these models can be classified beyond the linear theory. The dashed gray lines overlapped on the curves represent the torque value of the gas component. For all planet masses, the gas component dominates the total torque.
}
\label{fig:torq001}
\end{figure}
%%%%%%%%%%%%%%----------

For the dust component, we consider three different Stokes numbers $\mathrm{St}=\{10^{-3},0.01,0.1\}$. Even in a viscous disk, dust settles towards a scale height which tends to be lower than that of gas ($h_\mathrm{d}\neq h_\mathrm{g}$).
To initialize the dust density, we use Eq. (\ref{eq:rhod}) with
\begin{equation}
h_\mathrm{d}= h_\mathrm{g}\sqrt{\frac{\delta}{\delta+\mathrm{St}}},
    \label{eq:h_dust}
\end{equation}
which describes the equilibrium between dust settling and turbulent diffusion \citep[see][]{Dubrulle1995}. We set $\delta=10^{-3}$ for all Stokes number values.

Because the planet's mass is larger in this case, our models fall outside the linear regime (see below). Therefore, to adequately capture both thermal perturbations in the gas and the dust density structures, we perform global numerical simulations. Our computational domain extends for $0.48r_{p}$ to $2.08r_{p}$ in the radial direction, from $-\pi$ to $\pi$ in
the azimuthal direction, and between $\pi/2-2h_{p}$ to $\pi/2$ in colatitude. The number of zones in each direction is $(N_r,N_\phi,N_\theta)=(1864,1864,128)$. Once the domain of the computational box and the number of zones have been set, to increase the resolution around the planet, we use mesh density functions in both the radial and azimuthal directions\footnote{To create the high resolution mesh around the planet and a smooth transition with the coarse mesh, we use the coefficients $a_r=a_\phi=0.2618$, $b_r=b_\phi=0.3141$, $c_r=15.0$ and $c_\phi=1.5$, respectively.} \citep[see][]{BLl2023}, resulting in $\Delta r=\Delta\phi=5\times10^{-4}$. The computational domain entirely contains the 
gas and dust structures formed in the vicinity of the planet. Since we do not know a priori the effect of feedback on the gas structures in the vertical direction, we prefer to maintain a uniform mesh with a resolution similar to that used in \citet{ChM2021}. Note that the dust height scale is not sensitive to the change in vertical resolution \citep[see Appendix A in][]{CH2025}.

%%%%%%%%%%%%%%---------
\begin{figure}
\includegraphics[width=0.4853\textwidth]{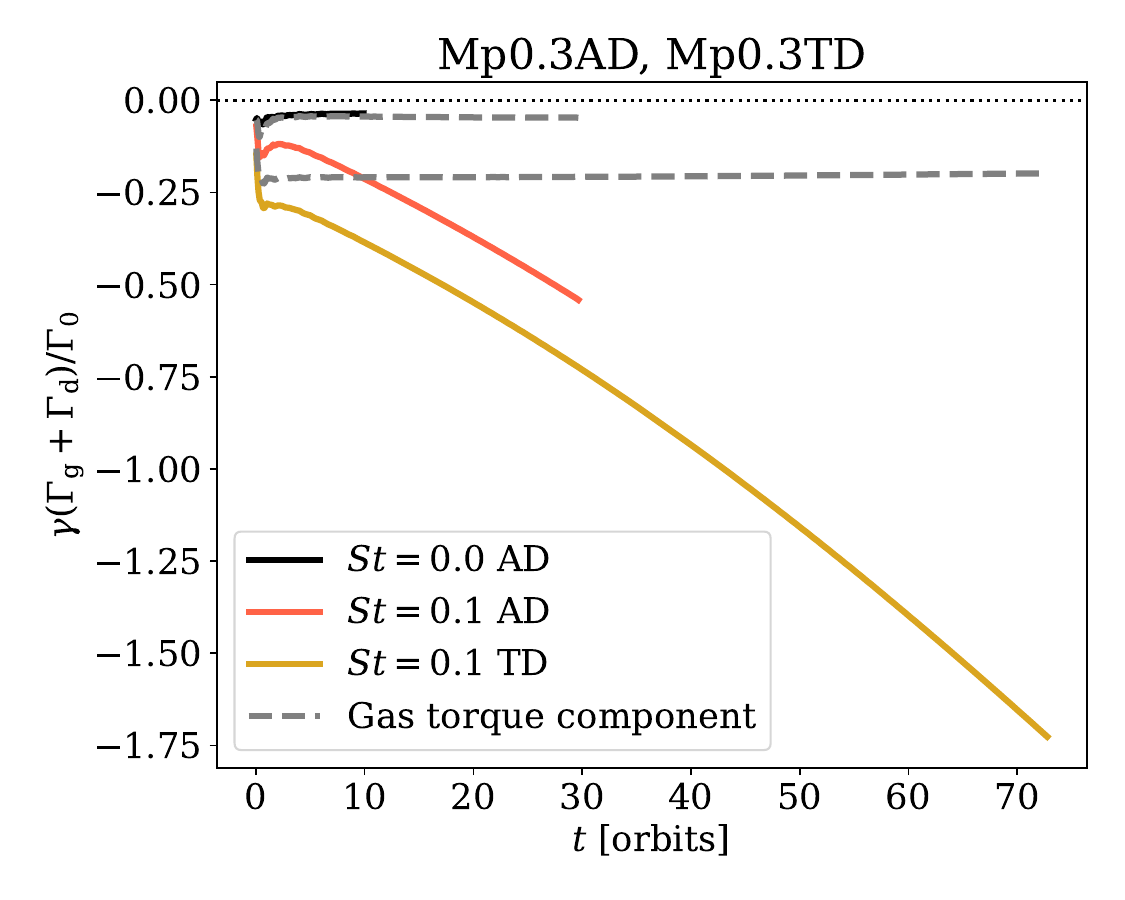}
 \caption{Normalized total torque (gas plus dust) as a function of time in the settling regime, for the global models Mp0.3AD and Mp0.3TD (see Table \ref{tab:t1}) considering that $\mathrm{St}=\{0.0,0.1\}$. The total torque is dominated by the dust torque.
}
\label{fig:torq1}
\end{figure}
%%%%%%%%%%%%%%----------

\subsection{Code and boundary conditions}

To solve numerically the continuity, momentum and gas-energy equations for the gas and dust, we use the hydrodynamic multi-fluid code \textsc{Fargo3D}\footnote{The public version of the code without the thermal diffusion module is available at \url{https://fargo3d.github.io/documentation/}.} \citep[][]{BLlM2016,BLlKP2019} with the fast orbital advection algorithm of \citet{Masset2000} and the rapid advection method \citep[RAM;][]{BLl2023}. 
A thermal diffusion module was implemented and tested in \citet{ChM2021} for a numerical study of thermal torques in dust-free disks. The RAM version of \textsc{Fargo3D} code allows us to increase the resolution sufficiently in the vicinity of the planet to capture the effects of thermal diffusion in the gas on a global grid of the protoplanetary disk.

We implemented boundary conditions for each dust species similar to those implemented for the density and velocity components of the gas 
\citep[for details, see][]{ChM2021}. In the radial direction, we use wave-damping boundary zones as in \citet{dVal2006};
the width of the inner and outer damping rings is at $r_i=0.915r_p$, $r_o=1.015r_p$, $r_i=0.65r_p$ and $r_o=1.65r_p$ for the local and global models, respectively.
The damping time-scale at the edge of each ring in both models is $0.3T_\mathrm{loc}$, with $T_\mathrm{loc}$ the local orbital period. At the upper meridional boundaries ($\theta_\mathrm{min}$),
we apply an extrapolation of the initial profiles for the gas and dust densities,
along with anti-symmetric boundary conditions for $v_\theta$ and $u_\theta$, respectively. At the lower meridional boundaries ($\theta_\mathrm{max}$), we use reflecting boundary conditions, since all our models include only one hemisphere of the disk.

%%%%%%%%%%%%%%---------
\begin{figure}
\includegraphics[width=0.4853\textwidth]{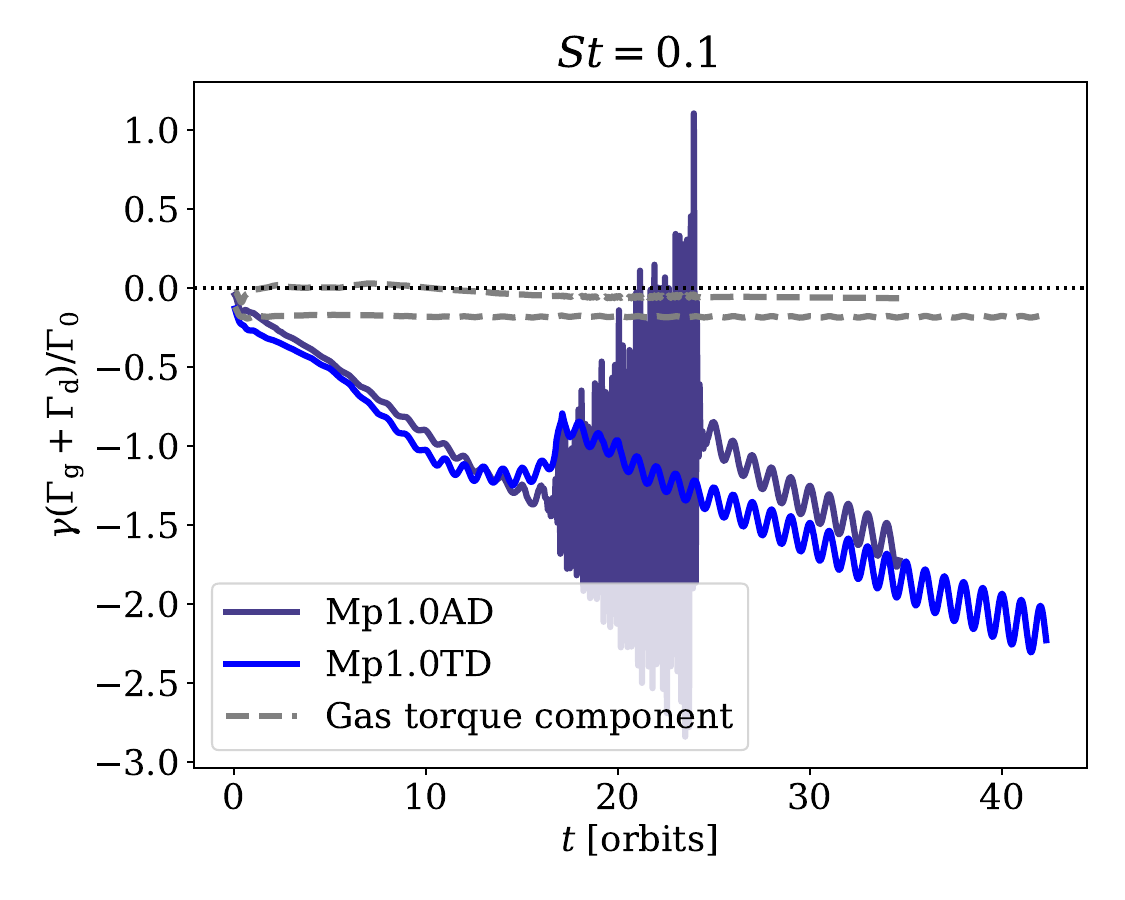}
 \caption{Similar to Fig. \ref{fig:torq1} but for a planetary mass of $M_p=1.0M_\oplus$ and $\mathrm{St}=0.1$. It can clearly be seen that the total torque is dominated by the dust torque component.
}
\label{fig:torq1Mp1}
\end{figure}
%%%%%%%%%%%%%%----------

%%%%%%%%%%%%%%%%%%%%

\section{Results}
\label{sec:results}

\subsection{Cold thermal torque within linear regime in gaseous-dusty disks}

In order to properly model dust settling, or the lack thereof, we start our local models with the dust scale height being equal to the gas scale height, that is, assuming that dust grains are initially well mixed with the gas.

\subsubsection{Cold thermal torque in a purely gaseous disk}

The cold thermal torque can be estimated in
linear theory if $x_p\lesssim \lambda_c \ll H_\mathrm{g}$
and the planetary mass is smaller than the critical mass
defined as:
\begin{equation}
M_c=\left(\frac{\lambda_c}{H_\mathrm{g}}\right)^2M_{\mathrm{th}},
    \label{eq:Mc}
\end{equation}
where $M_{\mathrm{th}}\equiv c_\mathrm{s}^3/(G\Omega_p)$ is the thermal mass \citep[see Sec. 6.1 in][]{ChM2021}.
If these conditions are fulfilled, the cold thermal
torque is given by 
\begin{equation}
\Gamma_{\mathrm{thermal}}^{\mathrm{cold}}=-1.61\frac{\gamma-1}{\gamma}\frac{x_p}{\lambda_c}\Gamma_0
    \label{eq:coldtorque}
\end{equation}
\citep{Masset2017}
We express $\Gamma_{\mathrm{thermal}}^{\mathrm{cold}}$ as a multiple of
\begin{equation}
\Gamma_0=\Sigma r_p^4\Omega_p^2\left(\frac{M_p}{M_\star}\right)^2h_\mathrm{g}^{-3},
    \label{eq:g0}
\end{equation}
instead of the usual expression that scales as $h_\mathrm{g}^{-2}$ \citep[see][]{Masset2017}. Equation (\ref{eq:coldtorque}) provides the torque in the absence of dust and will be used as a reference value to compare when we include the dust back-reaction.

From Table \ref{tab:t1} and the parameters introduced in Section~\ref{sec:init}, only the simulation with a planet of mass $M_p=0.03M_\oplus$ (model Mp0.03LR) satisfies the conditions for the linear regime. Then, if we substitute the values of $x_p$ and $\lambda_c$ in Eq.~(\ref{eq:coldtorque}), the cold thermal torque is $\Gamma_{\mathrm{thermal}}^{\mathrm{cold}}\simeq-0.36\Gamma_0$ in this case.

To compare the predicted cold thermal torque with simulation results, we ran a simulation of a low-mass planet embedded in a purely gaseous disk (that is, Mp0.03LR model for a Stokes number of $\mathrm{St}=0$, see Table \ref{tab:t1}). The evolution of the total gas torque is shown in Fig. ~\ref{fig:torques}. The value predicted by Eq.~(\ref{eq:coldtorque}) agrees with the simulation result to within 2.8~\%.
This indicates that our numerical approximation adequately captures the effect of the cold thermal torque.  Figure~\ref{fig:torques} can be directly compared with Fig. 5 of \citet{ChM2021}, as the case presented there for a low-mass planet with luminosity $L=0$ is equivalent to our Mp0.03LR model with $\mathrm{St}=0$.

\subsubsection{Cold thermal torque in the presence of dust feedback}

Having determined the value of the cold thermal torque for a low-mass planet embedded in a gas disk without dust, we now investigate how this torque is affected when dust feedback on the gas is included.  To this end, we have included dust in the Mp0.03LR model with a Stokes number $\mathrm{St}\in[0,0.28]$. The results of these numerical experiments are shown in Fig. ~\ref{fig:torques}. 

We find that, for $\mathrm{St}=2.8\times 10^{-5}$, the dust is strongly coupled to the gas and hence the gas torque does not change with respect to the simulation without any dust species (in Fig. ~\ref{fig:torques} case with $\mathrm{St}=0.0$). For dust with $\mathrm{St}=2.8\times 10^{-3}$, the cold thermal torque is less negative than the gas torque in the model with $\mathrm{St}=0.0$ (about $15\%$ less). The angular momentum transfer by the dust to the gas is more significant when $\mathrm{St}=2.8\times 10^{-2}$; the torque presents a greater offset ($\sim 25\%$). Finally, when $\mathrm{St}=0.28$, the cold thermal torque changes considerably. It is about $60\%$ lower than the value obtained in the case without feedback (cases $\mathrm{St}=0$ and $\mathrm{St}=2.8\times10^{-5}$). Notably, for
$\mathrm{St}=0.28$ and without dust feedback, the torque also converges to the value given by Eq.~(\ref{eq:coldtorque}). 

Some authors have investigated how torques are affected when the planet releases heat \citep[e.g.,][]{ChM2021}. It is instructive to compare the torque resulting from dust feedback with that arising from heat release. By comparing
with Figure 5 in \citet[][]{ChM2021}, we find that a non-accreting planet subject to the dust back-reaction experiences a 
torque similar to that of a planet with the same mass and a luminosity close to the critical luminosity $L_c$.

%%%%%%%%%%%%%%---------
\begin{figure}
\includegraphics[width=0.4853\textwidth]{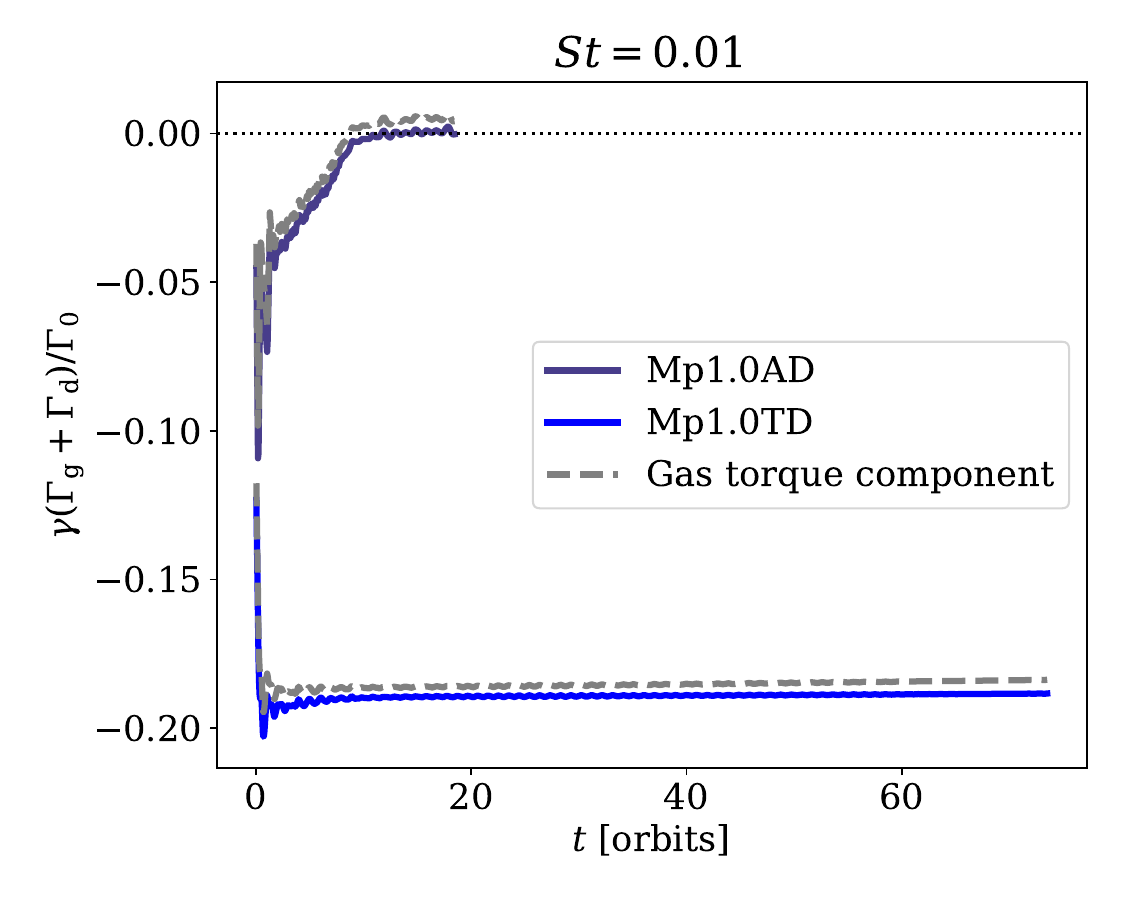}
 \caption{Same as Fig. \ref{fig:torq1Mp1} but for $\mathrm{St}=0.01$. In this case, we see that the total torque is dominated by the gas torque component.
}
\label{fig:torqS}
\end{figure}
%%%%%%%%%%%%%%----------

\begin{figure*}
    \centering
    \includegraphics[width=\linewidth]{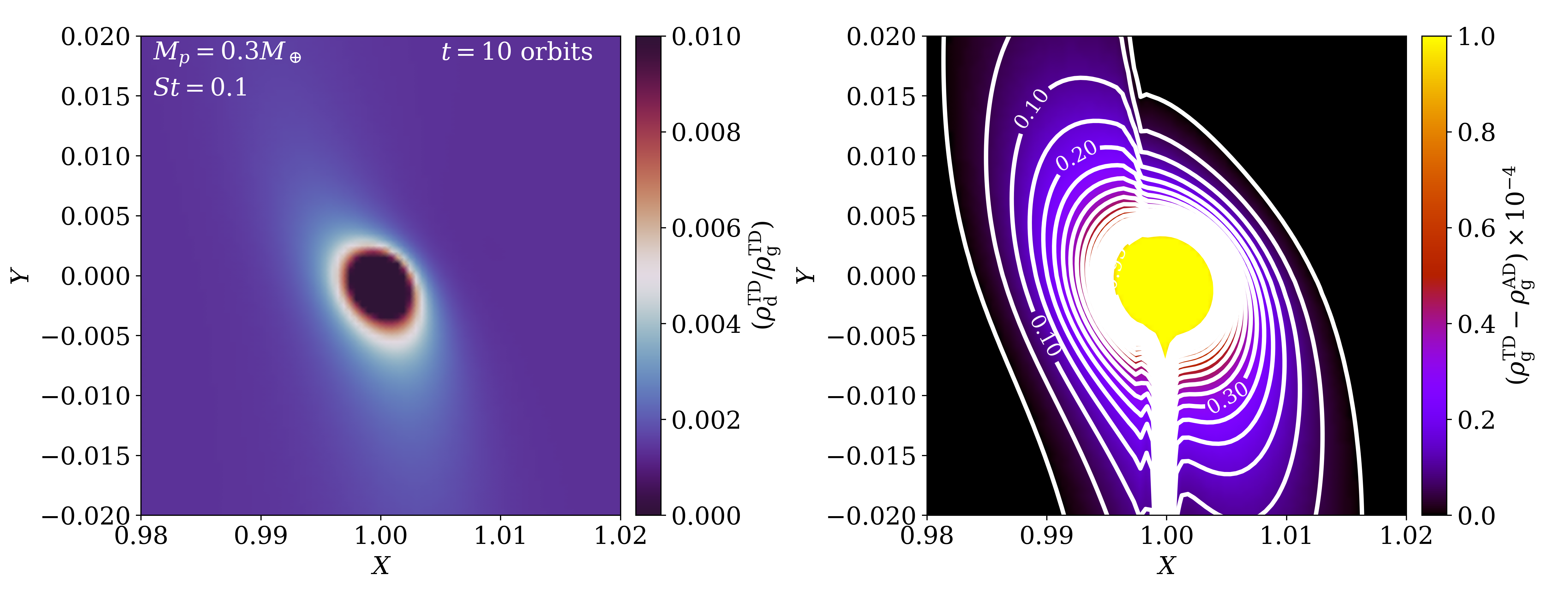}
    \caption{\textit{Left}. Perturbed gas density (code units) for the Mp0.3TD model with $\mathrm{St}=0.1$, at $t=10$ orbits. \textit{Right}. Cold thermal lobes in the gas component obtained from the subtraction of a simulation with thermal diffusivity (Mp0.3TD model) and an adiabatic simulation (Mp0.3AD model), at $t=10$ orbits. Both models have $\mathrm{St}=0.1$.}
    \label{fig:cl}
\end{figure*}

%%%%%%%%%%%%%%---------

\subsection{Beyond of the linear regime}

The results obtained in the previous section suggest that the effect of dust feedback on gas can be important on a low-mass planet, even without accreting pebbles. To investigate the effects of a more massive planet, we make use of the global disk models described in Section~\ref{subsec:global}. Since we are interested in analyzing the effect of different dust species on the total torque on the planet, we will describe the results in terms of the Stokes number.

Fig.~\ref{fig:torq001} shows the temporal evolution of the total torque and the gas torque in models Mp0.3AD, Mp0.3TD, Mp1.0AD and Mp1.0TD with $\mathrm{St}=0.001$. We find that, for the TD models, the total torque is more negative than for the AD models. We recall that in both the AD and TD models, we also included dust and its back-reaction force on the gas. Although the dust-to-gas ratio increases, then the dust feedback effect also increases, from Fig.~\ref{fig:torq001}, it is clear that the gas torque dominates in models with thermal diffusion,
i.e. the cold thermal torque.

Fig.~\ref{fig:torq1} shows the total torque and the gas torque on a planet with
$M_p=0.3$, for $\mathrm{St}=0.1$. While the gas torque remains approximately steady, the dust torque becomes increasingly negative. 
Within the first twenty orbital periods, the dust torque surpasses the gas torque in magnitude by a factor of two.

For a planet with $M_p=1.0M_\oplus$ and $\mathrm{St}=0.1$, the torque is again dominated by the dust component, which exhibits a decreasing oscillatory pattern (see Fig.~\ref{fig:torq1Mp1}). It should be mentioned that the difference in total torque between AD and TD models is significantly reduced as a result of the decreased coupling of the dust to the gas.

Particular attention should be paid to the case of a Stokes number of $\mathrm{St}=0.01$ in an adiabatic disk (see Fig.~\ref{fig:torqS}), since the gas torque can become positive. However, at the same time, the dust torque contributes just enough to such an extent that it counteracts the gas component, resulting in zero torque on the planet. For the Mp1.0TD model with $\mathrm{St}=0.01$, a similar behavior is observed, with the cold thermal torque prevailing over the dust torque. Lastly, we stress that the torque values presented have converged with respect to the numerical resolution \citep[see][]{CH2025}.

\section{Discussion}
\label{sec:discussion}

\subsection{Dust feedback freezes cold thermal torque}

As we showed in the previous section, it is clear that there is a bifurcation in the total torque value depending on the Stokes number. Here, let us first discuss the behavior of the gas component torque, which includes the cold thermal torque and the adiabatic case.

Remarkably, the value of the cold thermal torque appears to be insensitive to the Stokes number for a given value of the planet's mass. In other words, for both a planet with a mass of $M_p=0.3M_\oplus$ and a planet with a mass three times greater ($M_p=1M_\oplus$), and for all Stokes numbers explored, the value of the gas torque remains around $-0.2\gamma^{-1}\Gamma_0$ (see Figs.~\ref{fig:torq001}--\ref{fig:torqS}).

\begin{figure*}
    \centering
    \includegraphics[width=\linewidth]{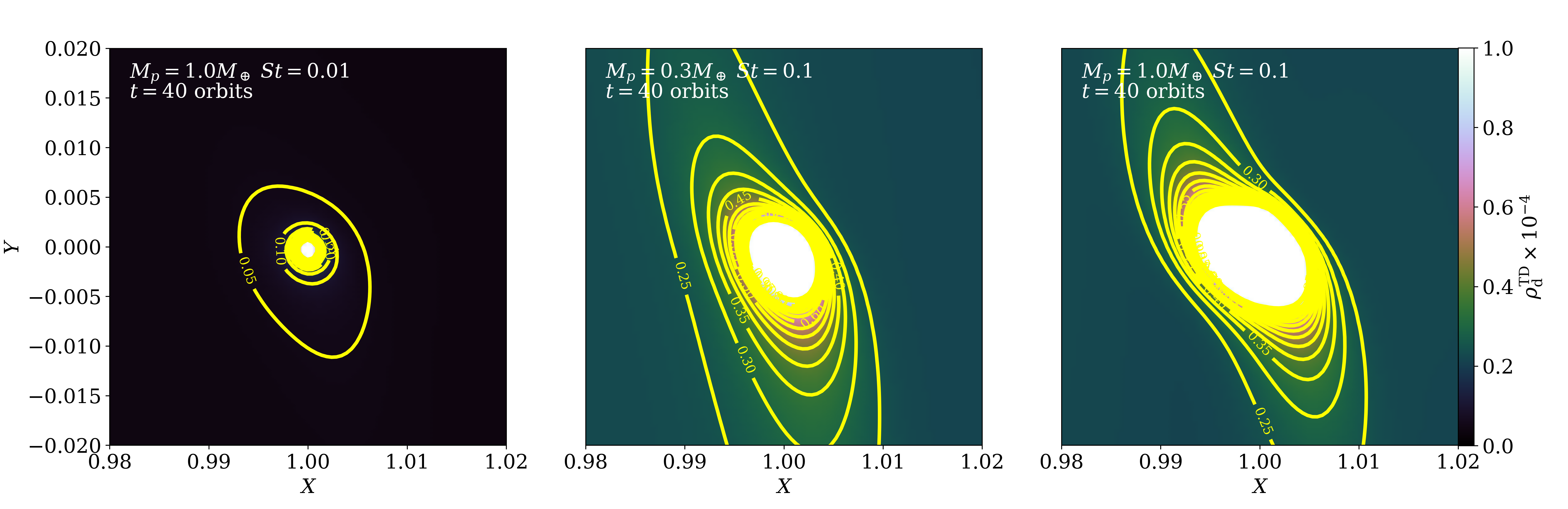}
    \caption{Dust density distribution around planets with mass $M_p=0.3M_\oplus$ and $M_p=1.0M_\oplus$ for Stokes numbers of $\mathrm{St}={0.01,0.1}$ (see text inside of plots and Table \ref{tab:t1}). It can be clearly seen that in the case of $\mathrm{St}=0.1$ there is the formation of dust lobes reminiscent of cold thermal lobes in the gas component.}
    \label{fig:dust_lobes}
\end{figure*}

This can be explained as follows. Thermal diffusivity produces perturbations of temperature and density of opposite signs without drastically changing the pressure \citep[][]{Masset2017}. For small Stokes numbers ($\mathrm{St}\leq10^{-2}$), dust tends to accumulate very close to the planet as a result of the greater coupling to the gas. For Stokes numbers
larger than $10^{-2}$, the coupling decreases and the gas maintains its structure generated by the diffusion-advection mechanism. Therefore, in both regimes the effects of turbulent diffusion and the dust feedback produce a counterweight to give rise to cold thermal lobes with an asymmetry to generate such a value of the torque.

Fig.~\ref{fig:cl} shows the general structure of the planetary wakes and the thermal lobes for the Mp0.3TD model with $\mathrm{St}=0.1$. The asymmetric overdensity around the planet (right panel in Fig.~\ref{fig:cl}) produces the cold thermal torque. Note that there is a filament emerging behind the planet position $(X,Y)=(1,0)$, this is an effect of the subtraction of the adiabatic simulation. In other words, it is a low-density region that only appears in the simulation without thermal diffusivity. Therefore, the shape of the cold thermal lobes is well defined as can be inferred from the behavior of the gas torque since no strong oscillations or peaks are observed once the steady state is reached.

In the case of the adiabatic torque, its dependence on the Stokes number is significant; for $\mathrm{St}=0.01$, the torque can be positive and,
in fact, reaches its maximum value compared to other Stokes numbers. As the Stokes number increases beyond this point, the torque decreases and eventually becomes negative.

\subsection{Dust streaming torque}

Recent 2D isothermal disk models neglecting dust feedback and dust turbulent diffusion \citep[][]{BLlP2018,Regaly2020} show that asymmetric dusty structures can be formed behind (underdense dust hole structure) or in front (overdense filamentary structure) of the planet. In many cases, the torque
generated by such structures is a positive quantity which can exceed in magnitude the gas torque \citep[see also][]{Chrenko2024}.

\citet{CH2025} showed that the dust density structures around a planet in an isothermal disk can evolve very differently over time compared to those in 2D models when dust feedback and turbulent diffusion are taken into account. Furthermore, the 3D simulations presented in \citet{CH2025} demonstrate that the overdense filament in front of the planet transforms into a trench, and that the dust hole (or void) can become deformed, potentially leading to either oscillatory or runaway migration depending on the strength of the feedback.

We emphasize that the models presented here represent both thermodynamic generalization, through the inclusion of adiabatic and thermally diffusive disk conditions, and an extension to a lower planetary mass regime ($M_p<1.5M_\oplus$) compared to the study by \citet{CH2025}.
This broader framework is particularly important, because 
decreasing the planet's mass and changing the gas thermodynamics reveals significant differences from those reported in that paper and in 
previous 2D studies. More specifically:
\begin{itemize}
    \item [(i)] In all global models where perturbations in dust density are correctly captured, regardless of whether they are adiabatic or with thermal diffusivity, and regardless of the value of the Stokes number, we find a negative dust torque. 
    \item [(ii)] When the dust torque dominates (that is, for $\mathrm{St}>10^{-2}$), we find the formation of high-density dust lobes around the planet (see Fig.~\ref{fig:dust_lobes}), which are reminiscent of the cold thermal lobes formed in the gas (henceforth called dust lobes). These dust lobes produce a dust torque that becomes increasingly negative over time, leading to an inward runaway migration.
\end{itemize}

The left panel of Fig.~\ref{fig:dust_lobes} displays the dust density in the midplane of the disk near a planet of mass $M_p=1.0M_\oplus$ for $\mathrm{St}=0.01$ (i.e. for model Mp1.0TD) at $t=40$ orbits. It is clearly seen that neither a dust void nor a trench forms in the planet's vicinity. Instead, a very small, quasi-symmetric dust lobe forms, which does not produce a torque exceeding that of the gas. Moreover, in both disk models (either adiabatic or those including  thermal diffusivity), we do not observe the formation of a dust void behind the planet or a trench in front of it. This is consistent with \citet{CH2025} who report that large-scale dust structures \citep{BLlP2018} are erased if the turbulent dust diffusion is sufficiently strong (note that our global scale simulations use a relatively large kinematic viscosity (Sec \ref{subsec:global}) and compare to Appendix \ref{ap:appendixA}).

\subsection{Further improvements in total torque measurement in gaseous-dusty disk models}

For the first time, we have analyzed here the effect of cold thermal torque and dust streaming torque on a planet in global disk models. Although this study represents a step beyond what has been done so far in three-dimensional disk models that include two fluids (gas and dust) and a more realistic thermodynamic treatment than an isothermal disk, some outstanding questions need to be addressed. Specifically, what happens when the planet is allowed to accrete dust and release energy? That is, the counterpart of cold thermal torque, such as heating torque \citep[see][]{BLl_etal2015,Masset2017,ChM2021}.
Also, what is the effect of including a dust size distribution instead of a constant Stokes number in the torque bifurcation reported here? Or what happens in the transition regions of the gas disk (for instance, the inner rim) where dust can accumulate and where thermal diffusivity can increase/decrease?

For instance, we adopt a constant thermal diffusivity of $\chi = 4.5 \times 10^{-5} r_p^2 \Omega_p$ in our simulations, in line with previous studies \citep[e.g.,][]{Lega_etal2014,VRM2020} and consistent with the value used in our earlier works. This choice is motivated by the need to enable the formation of cold lobes around the planet, a key ingredient for generating thermal torques in non-luminous cases. Maintaining a constant value of $\chi$ allows us to construct a thermally controlled environment, which is essential for isolating the effects of dust feedback on the gas structure and the resulting torque balance near the planet, the main focus of this study.

Nevertheless, we acknowledge that this is a simplification. In realistic protoplanetary disks, the thermal diffusivity $\chi$ is expected to vary spatially as it depends on local physical conditions, notably the gas density, temperature, and the opacity of the medium \citep[e.g.,][]{K2009,JM2017}. Specifically, $\chi$ scales approximately as $\chi \propto T^3 / \kappa$, where $\kappa$ is the mean opacity, and $T$ is the local temperature. Both of these quantities are, in turn, influenced by the local dust content, which depends on the size distribution, spatial concentration, and composition of the grains.

Our simulations show that the dust feedback onto the gas can significantly reshape the gas structure near the planet, particularly inside the cold lobes. These changes may, in principle, affect the distribution of dust, hence modifying the local temperature and opacity and, by extension, the thermal diffusivity. This leads to the intriguing possibility that $\chi$ itself may become asymmetric around the planet. Such an asymmetry could further distort the geometry of the cold lobes around the planet, increasing the asymmetry between the inner and outer contributions to the thermal torque, and thereby amplifying the net torque acting on the planet.

Exploring this effect in a self-consistent way would require a more complex, iterative modeling approach. First, one would need to simulate the coupled evolution of gas and dust, then calculate the local opacity from the dust distribution, update the temperature field and thermal diffusivity accordingly, and feed the resulting thermodynamic structure back into the dynamical evolution. This cycle would likely need to be repeated until a consistent steady state is reached. Additionally, this effort would require a more sophisticated treatment of dust opacity, including its dependence on composition, porosity, and mixing ratios, all of which remain highly uncertain and model-dependent.

Although this is undoubtedly a physically rich and promising problem — with potential implications for planetary migration, pebble accretion, and dust trapping — it falls beyond the scope of the present study. Here, our goal is to establish a first step in understanding how dust back-reaction interacts with thermal forces in a controlled setup. The consideration of spatially varying thermal diffusivity due to evolving dust distributions will be an exciting direction for future investigations.

We also mention that, in this study we have ruled out the inclusion of the effects of viscous heating, although we are aware that in the case of $\nu=10^{-5}r_p^2\Omega_p$ it could have some possible effect on the gas dynamics, the dominant torque is that of dust. Lastly, we stress that, due to the high computational cost of our models, answering these questions is beyond the scope of this work. However, these questions definitely deserve to be addressed in future research.

\section{Conclusions}
\label{sec:conclusions}
We conducted high-resolution, three-dimensional, two-fluid simulations of a low-mass planet embedded in a disk of gas and dust,
employing both local and global setups. In order to analyze the cold thermal torque \citep[][]{Lega_etal2014,Masset2017} under the influence of the aerodynamic back-reaction of different dust species.

Within the framework of linear theory using local mesh models for a low-mass planet ($M_p=0.03M_\oplus$). We find that with increasing Stokes number, the dust feedback reduces the magnitude of the linear cold thermal torque and the gas torque thus progressively shifts closer to the adiabatic limit (see Fig. ~\ref{fig:torques}).

In global disk models, we have resolved the cold thermal lobes in the non-linear regime and also captured the dust density structures responsible for the dust-driven torque. In these models, we find that regardless of the planet's mass, there is a bifurcation determined by the Stokes number $\mathrm{St}=0.01$, clearly identifying when each torque component dominates. For $\mathrm{St}\leq0.01$, the cold thermal torque dominates, since the formation of asymmetric substructures in the dust component that could contribute to the total torque on the planet is not observed. For $\mathrm{St}> 0.01$, we find\footnote{In fact, for the threshold value of the Stokes number above which dust toque becomes important, it can be readily estimated: in order for dust torque to become comparable to gas torque, their densities should be comparable at the midplane, that is $\rho_d/\rho_g\sim\epsilon h_g/h_d\sim 1$. From Eq. (\ref{eq:h_dust}), $\epsilon \sim (1+ \mathrm{St}/\delta)^{-1/2}$. Then the required value of $\mathrm{St}$ is: $\mathrm{St}\sim \epsilon^{-2} \delta \sim 0.1$, which is in agreement with our numerical results} the formation of overdense dust lobes with an asymmetry that results in a negative total torque that becomes more negative over time. These lobes in the dust are reminiscent of the cold thermal lobes formed in the gas and are the result of considering a non-isothermal disk. The outcome suggests that the formation of dust lobes could produce an inward migration or even trigger a runaway migration similar to \citet[][]{CH2025}.

Our results show, for the first time, under which conditions each of the torque components controls the planet dynamics. An important implication is that, due to the bifurcation in the total torque at $\mathrm{St}=0.01$, if the planet reaches its pebble isolation mass \citep{Lamb2014,Bert2018,Ch_etal2022}, the migration would be governed by the cold thermal torque, since the pebble isolation mass allows only dust grains with St $\lesssim$ 0.01 to enter the planet's vicinity. On the other hand, in disk regions with large dust grains and an increased dust-to-gas ratio (as in pressure bumps), the planet's dynamics should be governed by the dust lobes, i.e., the streaming torque.
  
\begin{acknowledgements}
     This work was supported by the Czech Science Foundation
    (grant 21-23067M). The work of O.C.\ was supported by the Charles University Research Centre program (No.~UNCE/24/SCI/005). Computational resources were available thanks to the Ministry of Education, Youth and Sports of the Czech Republic through the e-INFRA CZ (ID:~90254). JD and SC were funded by the European Union under the European Union’s Horizon Europe Research \& Innovation Programme 101040037 (PLANETOIDS).
    GDM acknowledges the support from the European Research Council under the Horizon 2020 Framework Program via the ERC Advanced Grant ``Origins'' (PI: Henning), Nr.~832428, and under the European Union's Horizon 2020 research and innovation programme ``PROTOPLANETS'' (PI: M.~Benisty), Nr.~101002188.
    Views and opinions expressed are however those of the authors only and do not necessarily reflect those of the European Union or the European Research Council. Neither the European Union nor the granting authority can be held responsible for them. YH is supported by the Jet Propulsion Laboratory, California Institute of Technology, under a contract with the National Aeronautics and Space Administration (80NM0018D0004).

\end{acknowledgements}

\bibliographystyle{yahapj}  % has links to the DOIs in the references
\bibliography{dustybib}

\begin{thebibliography}{}
\providecommand\natexlab[1]{#1}
\providecommand\JournalTitle[1]{#1}

\bibitem[{{Ben{\'\i}tez-Llambay} {et~al.}(2019){Ben{\'\i}tez-Llambay}, {Krapp}, \& {Pessah}}]{BLlKP2019}
{Ben{\'\i}tez-Llambay}, P., {Krapp}, L., \& {Pessah}, M.~E. 2019, \href{http://dx.doi.org/10.3847/1538-4365/ab0a0e}{\JournalTitle{\apjs}, 241, 25}

\bibitem[{{Ben{\'\i}tez-Llambay} {et~al.}(2023){Ben{\'\i}tez-Llambay}, {Krapp}, {Ramos}, \& {Kratter}}]{BLl2023}
{Ben{\'\i}tez-Llambay}, P., {Krapp}, L., {Ramos}, X.~S., \& {Kratter}, K.~M. 2023, \href{http://dx.doi.org/10.3847/1538-4357/acd698}{\JournalTitle{\apj}, 952, 106}

\bibitem[{{Ben{\'\i}tez-Llambay} {et~al.}(2015){Ben{\'\i}tez-Llambay}, {Masset}, {Koenigsberger}, \& {Szul{\'a}gyi}}]{BLl_etal2015}
{Ben{\'\i}tez-Llambay}, P., {Masset}, F., {Koenigsberger}, G., \& {Szul{\'a}gyi}, J. 2015, \href{http://dx.doi.org/10.1038/nature14277}{\JournalTitle{\nat}, 520, 63}

\bibitem[{{Ben{\'\i}tez-Llambay} \& {Masset}(2016)}]{BLlM2016}
{Ben{\'\i}tez-Llambay}, P., \& {Masset}, F.~S. 2016, \href{http://dx.doi.org/10.3847/0067-0049/223/1/11}{\JournalTitle{\apjs}, 223, 11}

\bibitem[{{Ben{\'\i}tez-Llambay} \& {Pessah}(2018)}]{BLlP2018}
{Ben{\'\i}tez-Llambay}, P., \& {Pessah}, M.~E. 2018, \href{http://dx.doi.org/10.3847/2041-8213/aab2ae}{\JournalTitle{\apjl}, 855, L28}

\bibitem[{{Bitsch} {et~al.}(2013){Bitsch}, {Boley}, \& {Kley}}]{BB2013}
{Bitsch}, B., {Boley}, A., \& {Kley}, W. 2013, \href{http://dx.doi.org/10.1051/0004-6361/201118490}{\JournalTitle{\aap}, 550, A52}

\bibitem[{{Bitsch} {et~al.}(2018){Bitsch}, {Morbidelli}, {Johansen}, {Lega}, {Lambrechts}, \& {Crida}}]{Bert2018}
{Bitsch}, B., {Morbidelli}, A., {Johansen}, A., {et~al.} 2018, \href{http://dx.doi.org/10.1051/0004-6361/201731931}{\JournalTitle{\aap}, 612, A30}

\bibitem[{{Bitsch} {et~al.}(2014){Bitsch}, {Morbidelli}, {Lega}, \& {Crida}}]{Bitsch2014}
{Bitsch}, B., {Morbidelli}, A., {Lega}, E., \& {Crida}, A. 2014, \href{http://dx.doi.org/10.1051/0004-6361/201323007}{\JournalTitle{\aap}, 564, A135}

\bibitem[{{Chametla} {et~al.}(2025){Chametla}, {Chrenko}, {Masset}, {D'Angelo}, \& {Nesvorn{\'y}}}]{CH2025}
{Chametla}, R.~O., {Chrenko}, O., {Masset}, F.~S., {D'Angelo}, G., \& {Nesvorn{\'y}}, D. 2025, \href{http://dx.doi.org/10.1051/0004-6361/202451869}{\JournalTitle{\aap}, 698, A21}

\bibitem[{{Chametla} {et~al.}(2024){Chametla}, {Chrenko}, {Reyes-Ruiz}, \& {S{\'a}nchez-Salcedo}}]{CH2024}
{Chametla}, R.~O., {Chrenko}, O., {Reyes-Ruiz}, M., \& {S{\'a}nchez-Salcedo}, F.~J. 2024, \href{http://dx.doi.org/10.1093/mnras/stad3898}{\JournalTitle{\mnras}, 527, 11812}

\bibitem[{{Chametla} \& {Masset}(2021)}]{ChM2021}
{Chametla}, R.~O., \& {Masset}, F.~S. 2021, \href{http://dx.doi.org/10.1093/mnras/staa3681}{\JournalTitle{\mnras}, 501, 24}

\bibitem[{{Chametla} {et~al.}(2022){Chametla}, {Masset}, {Baruteau}, \& {Bitsch}}]{Ch_etal2022}
{Chametla}, R.~O., {Masset}, F.~S., {Baruteau}, C., \& {Bitsch}, B. 2022, \href{http://dx.doi.org/10.1093/mnras/stab3753}{\JournalTitle{\mnras}, 510, 3867}

\bibitem[{{Chrenko} {et~al.}(2024){Chrenko}, {Chametla}, {Masset}, {Baruteau}, \& {Bro{\v{z}}}}]{Chrenko2024}
{Chrenko}, O., {Chametla}, R.~O., {Masset}, F.~S., {Baruteau}, C., \& {Bro{\v{z}}}, M. 2024, \href{http://dx.doi.org/10.1051/0004-6361/202450922}{\JournalTitle{\aap}, 690, A41}

\bibitem[{{Crida} {et~al.}(2022){Crida}, {Griveaud}, {Lega}, {Masset}, {Morbidelli}, {Kloster}, {Marques}, \& {Minker}}]{Cri2022}
{Crida}, A., {Griveaud}, P., {Lega}, E., {et~al.} 2022, in SF2A-2022: Proceedings of the Annual meeting of the French Society of Astronomy and Astrophysics, ed. J.~{Richard}, A.~{Siebert}, E.~{Lagadec}, N.~{Lagarde}, O.~{Venot}, J.~{Malzac}, J.~B. {Marquette}, M.~{N'Diaye}, \& B.~{Briot}, 315

\bibitem[{{D'Angelo} \& {Bodenheimer}(2013)}]{DAB13}
{D'Angelo}, G., \& {Bodenheimer}, P. 2013, \href{http://dx.doi.org/10.1088/0004-637X/778/1/77}{\JournalTitle{\apj}, 778, 77}

\bibitem[{{de Val-Borro} {et~al.}(2006){de Val-Borro}, {Edgar}, {Artymowicz}, {Ciecielag}, {Cresswell}, {D'Angelo}, {Delgado-Donate}, {Dirksen}, {Fromang}, {Gawryszczak}, {Klahr}, {Kley}, {Lyra}, {Masset}, {Mellema}, {Nelson}, {Paardekooper}, {Peplinski}, {Pierens}, {Plewa}, {Rice}, {Sch{\"a}fer}, \& {Speith}}]{dVal2006}
{de Val-Borro}, M., {Edgar}, R.~G., {Artymowicz}, P., {et~al.} 2006, \href{http://dx.doi.org/10.1111/j.1365-2966.2006.10488.x}{\JournalTitle{\mnras}, 370, 529}

\bibitem[{{Dubrulle} {et~al.}(1995){Dubrulle}, {Morfill}, \& {Sterzik}}]{Dubrulle1995}
{Dubrulle}, B., {Morfill}, G., \& {Sterzik}, M. 1995, \href{http://dx.doi.org/10.1006/icar.1995.1058}{\JournalTitle{\icarus}, 114, 237}

\bibitem[{{Eklund} \& {Masset}(2017)}]{EM2017}
{Eklund}, H., \& {Masset}, F.~S. 2017, \href{http://dx.doi.org/10.1093/mnras/stx856}{\JournalTitle{\mnras}, 469, 206}

\bibitem[{{Hasegawa} {et~al.}(2017){Hasegawa}, {Okuzumi}, {Flock}, \& {Turner}}]{Hasegawa2017}
{Hasegawa}, Y., {Okuzumi}, S., {Flock}, M., \& {Turner}, N.~J. 2017, \href{http://dx.doi.org/10.3847/1538-4357/aa7d55}{\JournalTitle{\apj}, 845, 31}

\bibitem[{{Jim{\'e}nez} \& {Masset}(2017)}]{JM2017}
{Jim{\'e}nez}, M.~A., \& {Masset}, F.~S. 2017, \href{http://dx.doi.org/10.1093/mnras/stx1946}{\JournalTitle{\mnras}, 471, 4917}

\bibitem[{{Kley} {et~al.}(2009){Kley}, {Bitsch}, \& {Klahr}}]{K2009}
{Kley}, W., {Bitsch}, B., \& {Klahr}, H. 2009, \href{http://dx.doi.org/10.1051/0004-6361/200912072}{\JournalTitle{\aap}, 506, 971}

\bibitem[{{Krapp} {et~al.}(2024){Krapp}, {Kratter}, {Youdin}, {Ben{\'\i}tez-Llambay}, {Masset}, \& {Armitage}}]{Krapp2024}
{Krapp}, L., {Kratter}, K.~M., {Youdin}, A.~N., {et~al.} 2024, \href{http://dx.doi.org/10.3847/1538-4357/ad644a}{\JournalTitle{\apj}, 973, 153}

\bibitem[{{Lambrechts} {et~al.}(2014){Lambrechts}, {Johansen}, \& {Morbidelli}}]{Lamb2014}
{Lambrechts}, M., {Johansen}, A., \& {Morbidelli}, A. 2014, \href{http://dx.doi.org/10.1051/0004-6361/201423814}{\JournalTitle{\aap}, 572, A35}

\bibitem[{{Lega} {et~al.}(2014){Lega}, {Crida}, {Bitsch}, \& {Morbidelli}}]{Lega_etal2014}
{Lega}, E., {Crida}, A., {Bitsch}, B., \& {Morbidelli}, A. 2014, \href{http://dx.doi.org/10.1093/mnras/stu304}{\JournalTitle{\mnras}, 440, 683}

\bibitem[{{Lesur} {et~al.}(2025){Lesur}, {Latter}, \& {Ogilvie}}]{Lesur2025}
{Lesur}, G., {Latter}, H.~N., \& {Ogilvie}, G.~I. 2025, \href{http://dx.doi.org/10.48550/arXiv.2508.20839}{\JournalTitle{arXiv e-prints}, arXiv:2508.20839}

\bibitem[{{Lesur} {et~al.}(2022){Lesur}, {Ercolano}, {Flock}, {Lin}, {Yang}, {Barranco}, {Benitez-Llambay}, {Goodman}, {Johansen}, {Klahr}, {Laibe}, {Lyra}, {Marcus}, {Nelson}, {Squire}, {Simon}, {Turner}, {Umurhan}, \& {Youdin}}]{Lesur_etal2022}
{Lesur}, G., {Ercolano}, B., {Flock}, M., {et~al.} 2022, \href{http://dx.doi.org/10.48550/arXiv.2203.09821}{\JournalTitle{arXiv e-prints}, arXiv:2203.09821}

\bibitem[{{Masset}(2000)}]{Masset2000}
{Masset}, F. 2000, \href{http://dx.doi.org/10.1051/aas:2000116}{\JournalTitle{\aaps}, 141, 165}

\bibitem[{{Masset}(2017)}]{Masset2017}
{Masset}, F.~S. 2017, \href{http://dx.doi.org/10.1093/mnras/stx2271}{\JournalTitle{\mnras}, 472, 4204}

\bibitem[{{Masset} \& {Ben{\'\i}tez-Llambay}(2016)}]{MBLl2016}
{Masset}, F.~S., \& {Ben{\'\i}tez-Llambay}, P. 2016, \href{http://dx.doi.org/10.3847/0004-637X/817/1/19}{\JournalTitle{\apj}, 817, 19}

\bibitem[{{Morfill} \& {Voelk}(1984)}]{MorfilV1984}
{Morfill}, G.~E., \& {Voelk}, H.~J. 1984, \href{http://dx.doi.org/10.1086/162697}{\JournalTitle{\apj}, 287, 371}

\bibitem[{{M{\"u}ller} {et~al.}(2012){M{\"u}ller}, {Kley}, \& {Meru}}]{Muller2012}
{M{\"u}ller}, T.~W.~A., {Kley}, W., \& {Meru}, F. 2012, \href{http://dx.doi.org/10.1051/0004-6361/201118737}{\JournalTitle{\aap}, 541, A123}

\bibitem[{{Murray} \& {Dermott}(1999)}]{MD1999}
{Murray}, C.~D., \& {Dermott}, S.~F. 1999, {Solar System Dynamics}

\bibitem[{{Ogilvie} {et~al.}(2025){Ogilvie}, {Latter}, \& {Lesur}}]{Ogil2025}
{Ogilvie}, G.~I., {Latter}, H.~N., \& {Lesur}, G. 2025, \href{http://dx.doi.org/10.1093/mnras/staf154}{\JournalTitle{\mnras}, 537, 3349}

\bibitem[{{Paardekooper} {et~al.}(2023){Paardekooper}, {Dong}, {Duffell}, {Fung}, {Masset}, {Ogilvie}, \& {Tanaka}}]{Paar_etal2023}
{Paardekooper}, S., {Dong}, R., {Duffell}, P., {et~al.} 2023, \href{http://dx.doi.org/10.48550/arXiv.2203.09595}{in Astronomical Society of the Pacific Conference Series, Vol. 534, Protostars and Planets VII, ed. S.~{Inutsuka}, Y.~{Aikawa}, T.~{Muto}, K.~{Tomida}, \& M.~{Tamura}}, 685

\bibitem[{{Paardekooper} {et~al.}(2011){Paardekooper}, {Baruteau}, \& {Kley}}]{Paar2011}
{Paardekooper}, S.~J., {Baruteau}, C., \& {Kley}, W. 2011, \href{http://dx.doi.org/10.1111/j.1365-2966.2010.17442.x}{\JournalTitle{\mnras}, 410, 293}

\bibitem[{{Reg{\'a}ly}(2020)}]{Regaly2020}
{Reg{\'a}ly}, Z. 2020, \href{http://dx.doi.org/10.1093/mnras/staa2181}{\JournalTitle{\mnras}, 497, 5540}

\bibitem[{{Shakura} \& {Sunyaev}(1973)}]{SS1973}
{Shakura}, N.~I., \& {Sunyaev}, R.~A. 1973, \JournalTitle{\aap}, 24, 337

\bibitem[{{Tanaka} {et~al.}(2002){Tanaka}, {Takeuchi}, \& {Ward}}]{TTW2002}
{Tanaka}, H., {Takeuchi}, T., \& {Ward}, W.~R. 2002, \href{http://dx.doi.org/10.1086/324713}{\JournalTitle{\apj}, 565, 1257}

\bibitem[{{Vaidya} {et~al.}(2015){Vaidya}, {Mignone}, {Bodo}, \& {Massaglia}}]{Vetal2015}
{Vaidya}, B., {Mignone}, A., {Bodo}, G., \& {Massaglia}, S. 2015, \href{http://dx.doi.org/10.1051/0004-6361/201526247}{\JournalTitle{\aap}, 580, A110}

\bibitem[{{Velasco Romero} \& {Masset}(2020)}]{VRM2020}
{Velasco Romero}, D.~A., \& {Masset}, F.~S. 2020, \href{http://dx.doi.org/10.1093/mnras/staa1215}{\JournalTitle{\mnras}, 495, 2063}

\bibitem[{{Velasco Romero} {et~al.}(2022){Velasco Romero}, {Masset}, \& {Teyssier}}]{VRetal2022}
{Velasco Romero}, D.~A., {Masset}, F.~S., \& {Teyssier}, R. 2022, \href{http://dx.doi.org/10.1093/mnras/stab3334}{\JournalTitle{\mnras}, 509, 5622}

\bibitem[{{Ward}(1986)}]{Ward1986}
{Ward}, W.~R. 1986, \href{http://dx.doi.org/10.1016/0019-1035(86)90182-X}{\JournalTitle{\icarus}, 67, 164}

\bibitem[{{Weber} {et~al.}(2018){Weber}, {Ben{\'\i}tez-Llambay}, {Gressel}, {Krapp}, \& {Pessah}}]{Webber_etal2018}
{Weber}, P., {Ben{\'\i}tez-Llambay}, P., {Gressel}, O., {Krapp}, L., \& {Pessah}, M.~E. 2018, \href{http://dx.doi.org/10.3847/1538-4357/aaab63}{\JournalTitle{\apj}, 854, 153}

\bibitem[{{Zhu} {et~al.}(2015){Zhu}, {Stone}, \& {Bai}}]{Zhu2015}
{Zhu}, Z., {Stone}, J.~M., \& {Bai}, X.-N. 2015, \href{http://dx.doi.org/10.1088/0004-637X/801/2/81}{\JournalTitle{\apj}, 801, 81}

\end{thebibliography}


%%%%%%%%%%%%%%%%% APPENDICES %%%%%%%%%%%%%%%%%%%%%

\appendix
%\section{Increasing global dust-to-gas ratio}
%\label{ap:appendixA}

\section{Approximate isotropic low-turbulent diffusion}
\label{ap:appendixA}

\begin{figure}
\includegraphics[width=0.4853\textwidth]{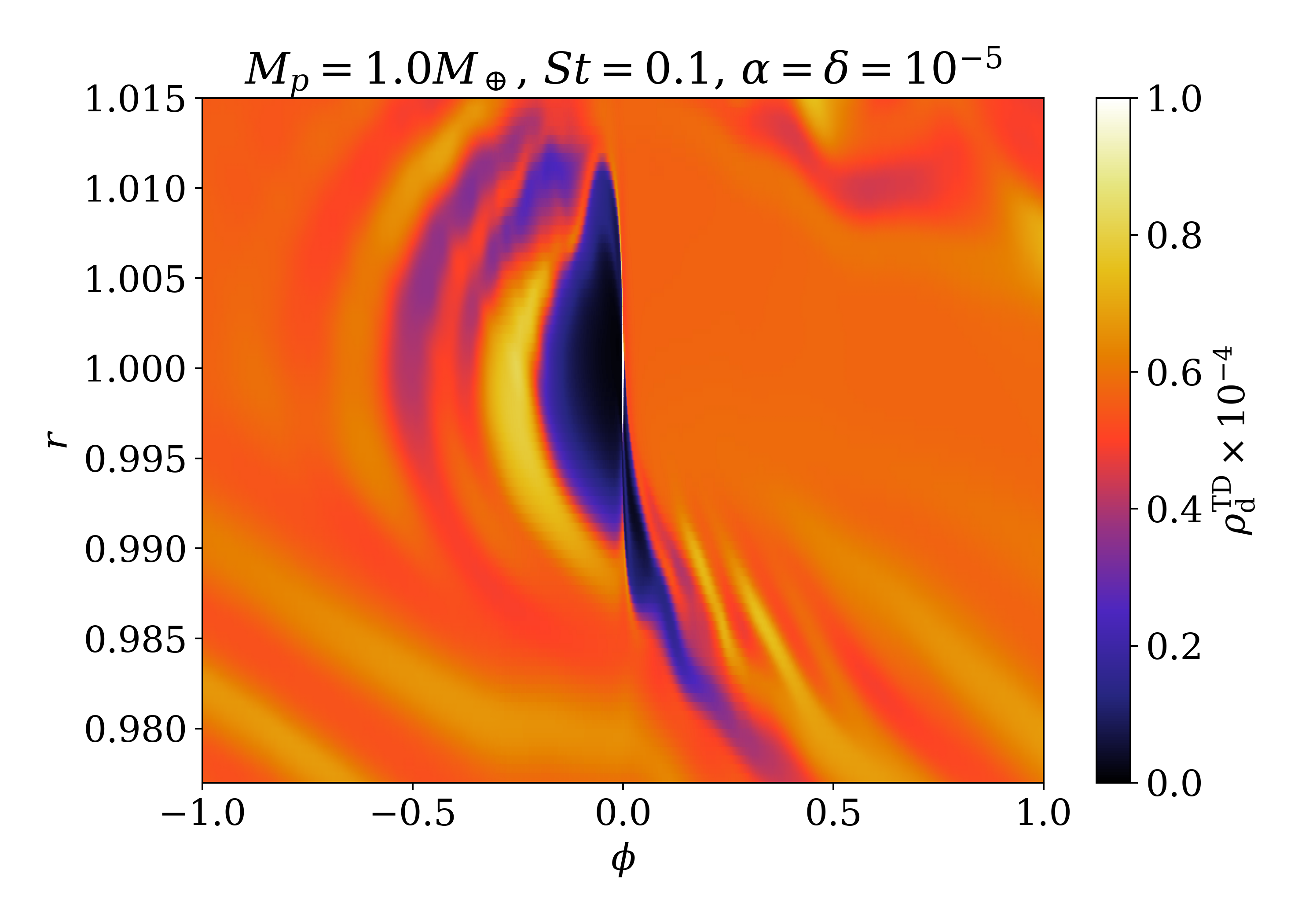}
 \caption{Dust-void structure at $t=10$ orbits, created by an Earth-like planet embedded in a protoplanetary disk with an isotropic turbulent diffusion.
}
\label{fig:tb}
\end{figure}

In the isotropic approximation of turbulent diffusion, the gas turbulence parameterized by alpha is assumed to be of the same magnitude as the dust turbulent diffusion. \citet{CH2025} recently studied the dust distribution around an Earth-like planet embedded in an isothermal protoplanetary disk. They found that in the low-turbulent regime $\delta=\alpha=10^{-5}$, a dust void forms behind the planet and a trench in front of it. The dust void is partially delimited by a high-density and a low-density filament (see their figure 4). Here, we similarly find the formation of a dust void and a trench, as well as a pair of low- and high-density structures around them (see Fig. \ref{fig:tb}).

%%%%%%%%%%%%%%---------
%%%%%%%%%%%%%%----------

\section{Smoothing length impact in 3D global models}
\label{ap:appendixB}

\begin{figure}
\includegraphics[width=0.4853\textwidth]{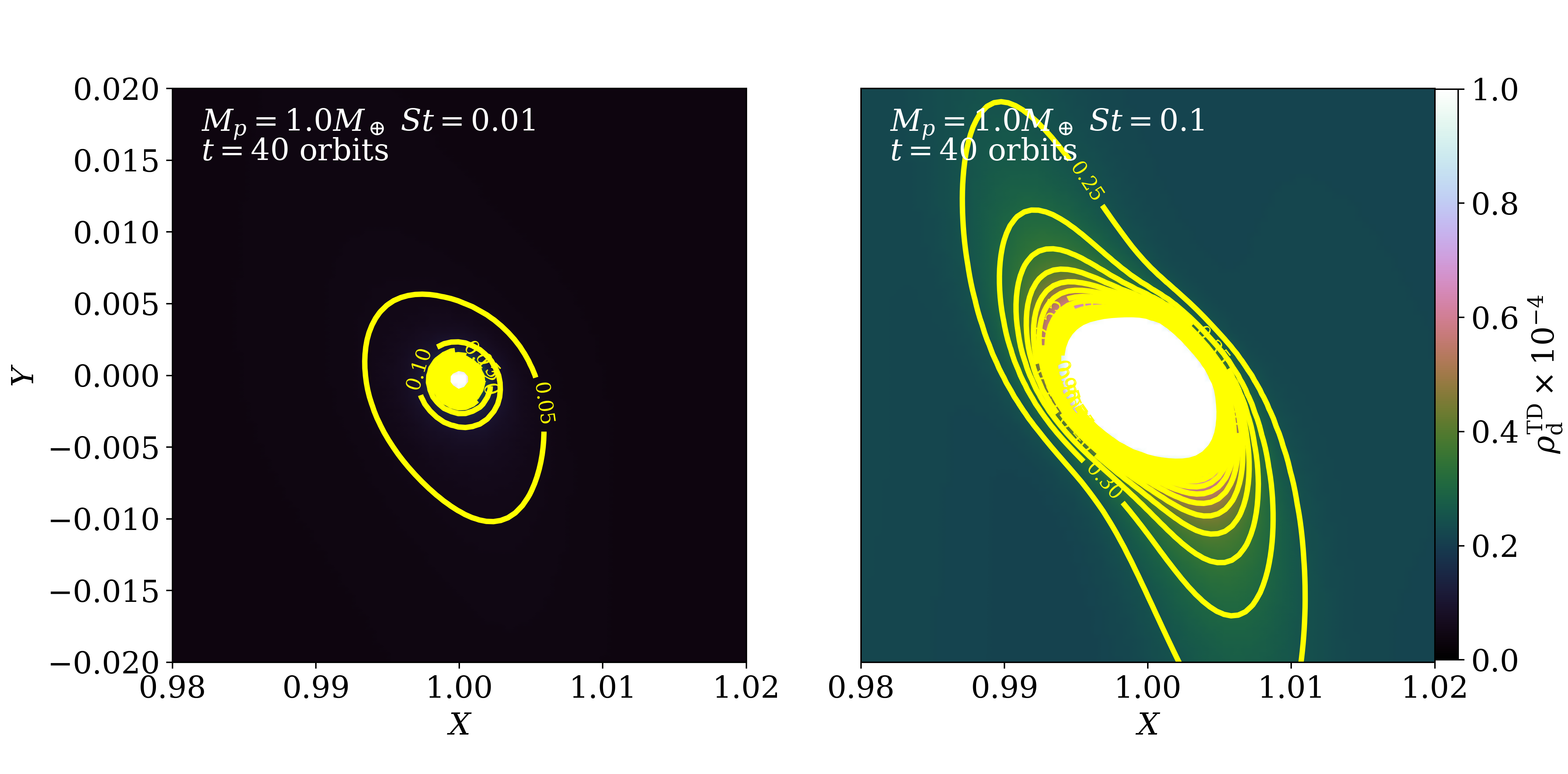}
 \caption{Dust density distribution in global models around of a planet with $M_p=1.0M_\oplus$ and Stokes numbers of $\mathrm{St}={0.01}$ and ${0.1}$ for the case of larger smoothing length than the one used in the main text.
}
\label{fig:sm}
\end{figure}

In Fig. \ref{fig:sm} we show the dust density distribution around a planet with $M_p=1.0M_\oplus$ for Stokes numbers of $\mathrm{St}=0.01$ (left panel) and $\mathrm{St}=0.1$ (right panel) considering a smoothing length of $r_\mathrm{sm}=0.03H_\mathrm{g}$, three times larger than that of our global simulations. We find that, for these values of the Stokes number, $r_\mathrm{sm}$ has only a weak impact on the gas and dust densities, and hence the total torque on the planet. For $\mathrm{St}=0.01$ the change in dust density was less than $2\%$, and for $\mathrm{St}=0.1$ it was less than $5\%$ (see Fig. \ref{fig:dust_lobes} in the main text). Interestingly, we find a small decrease in the dust torque in both cases, similar to what occurs in 2D simulations \citep[see][]{Regaly2020}. In a forthcoming paper (Chametla et al. in prep.), this effect of the smoothing length and the accretion of gas (heating torque component) and dust by the planet will be addressed since, as recently shown in 2D models \citep[][]{Chrenko2024}, the accretion of pebbles can considerably change the torque on the planet.

%%%%%%%%%%%%%%---------
%%%%%%%%%%%%%%----------

\end{document}